\documentclass[aps,prx,twocolumn]{revtex4}
\usepackage{graphicx}

\begin{document}

\title{Anisotropic Electronic Mobilities in the Nematic State of the Parent Phase NaFeAs}

\author{Qiang Deng, Jie Xing, Jianzhong Liu, Huan Yang$^*$, and Hai-Hu Wen}\email{huanyang@nju.edu.cn, hhwen@nju.edu.cn}

\affiliation{Center for Superconducting Physics and
Materials, National Laboratory of Solid State Microstructures and
Department of Physics, Collaborative Innovation Center of Advanced Microstructures, Nanjing University, Nanjing 210093, China}

\begin{abstract}
Hall effect and magnetoresistance have been measured on single crystals of the parent phase NaFeAs under a uniaxial pressure. Although significant difference of the in-plane resistivity $\rho_{xx}(I\parallel a)$ and $\rho_{xx}(I\parallel b)$ with the uniaxial pressure along $b$-axis was observed, the transverse resistivity $\rho_{xy}$ shows a surprisingly isotropic behavior. Detailed analysis reveals that the Hall coefficient $R_\mathrm{H}$ measured in the two orthogonal configurations ($I\parallel a$-axis and $I\parallel b$-axis) coincide very well and exhibit a deviation from the high temperature background at around the structural transition temperature $T_{\mathrm{s}}$. Furthermore, the magnitude of $R_\mathrm{H}$ increases remarkably below the structural transition temperature. This enhanced Hall coefficient is accompanied by the non-linear transverse resistivity versus magnetic field and enhanced magnetoresistance, which can be explained very well by the two band model with anisotropic mobilities of each band. Our results together with the two band model analysis clearly show that the anisotropic in-plane resistivity in the nematic state is closely related to the distinct quasiparticle mobilities when they are moving parallel or perpendicular to the direction of the uniaxial pressure.

\end{abstract}


\maketitle

\section{introduction}
The multiband nature of iron based superconductors make it complex and charming\cite{Singh,ChangLiu,HDing}. In many iron based superconductors, a nematic electronic state has been observed or suggested in the normal state through measurements of scanning tunneling spectroscopy (STS)\cite{DavisScience,Pasupathy,WangYY}, inelastic neutron scattering\cite{Dhital,PCDai1,PCDai2}, magnetic torque\cite{TorqueMatsuda}, point contact tunneling\cite{LGreene}, etc. The nematic state, by its definition, should have a $C_2$ symmetry of electronic property, which has been indeed observed directly in the STS measurements. Usually the normal state has a tetragonal structure at a high temperature, it changes to an orthorhombic phase at the structural transition temperature $T_{\text{s}}$. In the orthorhombic phase, the material will naturally form some twin boundaries, therefore the macroscopic probes, like resistivity would detect a global feature of the twined structure. However, if one applies a strain along one of the principal axis and the temperature is cooled down through $T_{\mathrm{s}}$, the material will be in a detwined state and the resistive measurement would be possible to reveal the nematic electronic state through the anisotropic resistivity. This interesting state was indeed detected by the in-plane resistive measurements in the Co-doped BaFe$_2$As$_2$ (Ba122) phase\cite{Fisher1,Fisher2,Prozorov1,Ying1} and NaFe$_{1-x}$Co$_x$As (Na111) phase\cite{DengQPRB}. In addition, in the hole-doped Ba122 phase\cite{Ruslan} and Ca122 phase\cite{XHChen}, a sign reversal of in-plane resistivity anisotropy has been observed. Some theoretical models have been developed to interpret the nematic behavior\cite{Fernandes,Eremin,WeiKu}. The central issue is to answer what is the driving force of nematicity, spin fluctuations or charge/orbital fluctuations? For this purpose, a great deal of researches have been developed, including Raman \cite{Yann,Blumberg1,Blumberg2}, optical\cite{Dusza,Nakajima}, angle-resolved photoemission spectroscopy (ARPES)\cite{ZXShen,ARPES-Feng}, etc. However, as far as we know, there is no consensus yet about what is the fundamental mechanism of the nematic state. It is very curious to know whether the nematicity is related to the structural, antiferromagnetic (AF) transitions, or orbital fluctuations. By post-annealing the samples, it was found that the distinction of the in-plane anisotropic resistivity can be lowered down, which initiates the discussion that the nematic state may be related to the local impurity scattering\cite{Uchida,DavisNatPhys}. In this paper, we report the in-plane resistivity, Hall effect and magneto-resistivity measurements in the parent phase NaFeAs under a uniaxial pressure with two orthogonal configurations: $I\parallel a$-axis and $I\parallel b$-axis. Our results show an isotropic transverse resistivity and Hall coefficient, indicating that the significant in-plane anisotropic resistivity is purely coming from the distinct quasiparticle mobilities in the nematic state.

\section{Experimental techniques}

The NaFeAs single crystals were grown by flux method using NaAs as flux. The details of synthesis was given in our previous paper\cite{Deng}. In this study, the Hall effect and magneto-resistivity measurements were performed simultaneously on a Quantum Design instrument (PPMS) using a standard six-lead method. The detwinning device used in this work is the same as that used in our previous study\cite{DengQPRB}. As shown in the inset of Fig.~1 on the left-hand side, a NaFeAs single crystal with nearly a square shape (3.8$\times$3.6$\times$0.12 mm$^3$) is mounted on the detwinning device, and the device is insulated by covering a piece of insulating sheet. The pressure applied in this measurement was about 2.5 MPa (estimated from the deformation of the spring under pressure and the cross-sectional area of the sample). As we know, in the orthorhombic phase, $b$-axis naturally aligns in the direction of the applied uniaxial pressure. Thus, the inset shows a configuration of the current applied parallel to $a$-axis. The insets in Fig.~1 show a photo (left) and a schematic picture (right) of the measurement setup with the pressure applied along $b$-axis and $I\parallel a$-axis. Hall voltage was measured along $b$-axis in this case. The measurement of $I\parallel b$-axis was performed on the same sample with the electrodes rotated $90^{\circ}$.  The measuring current was 1 mA. The longitudinal and transverse resistivity were measured with sweeping magnetic field from -9 T to 9 T at a fixed temperature. During the measurements, the magnetic field was applied perpendicular to the $ab$-plane of the sample. The longitudinal resistivity $\rho_{xx}$ was calculated by the averaged value of the resistivity measured at the magnetic fields with the same magnitude but opposite directions, while the transverse resistivity $\rho_{xy}$ was calculated by the difference of the two corresponding values at positive and negative magnetic fields to reduce the offset voltage caused by the possible nonsymmetric electric contact.

\begin{figure}
\includegraphics[width=9cm]{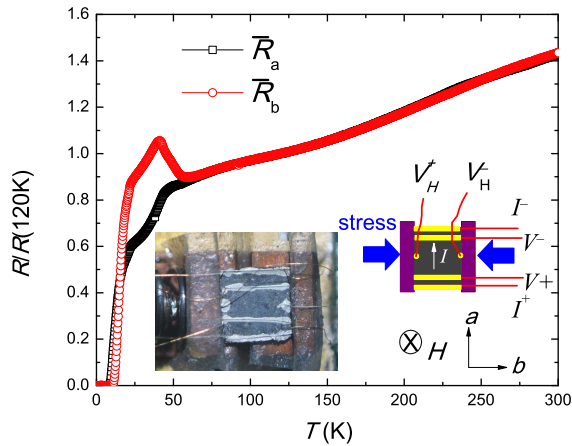}
\caption{Temperature dependence of normalized in-plane resistivity for one NaFeAs single crystal under a uniaxial pressure along $b$-axis. Here $\bar{R}_a$ and $\bar{R}_b$ represent the normalized resistance when the measuring current is along $a$-axis and $b$-axis respectively. Both resistance were normalized to the data at $T = 120$ K for comparison. The insets show the measurement setup with the pressure applied along $b$-axis and $I\parallel a$-axis. For the configuration with $I\parallel b$-axis (not shown here), we made the new electrodes on the same crystal and keep the pressure along $b$-axis.}
\end{figure}

\section{Results}

Figure~1 shows the temperature dependence of the in-plane resistivity for the NaFeAs single crystal under a uniaxial pressure along $b$-axis. $\bar{R}_a$ and $\bar{R}_b$ are the normalized resistance when the measuring current is along $a$-axis and $b$-axis, respectively. For a good comparison, both curves were normalized to the data at $T$ = 120 K. The kinky structures on the resistance curve are related to the structure and antiferromagnetic transitions. Following our previous method\cite{DengQPRB}, the transition temperatures $T_{\text{s}}\approx52$ K and $T_{\text{AF}}\approx43$ K are determined from the derivative curve of $\bar{R}_b$. A clear distinction between $\bar{R}_a$ and $\bar{R}_b$ can be observed in the low-temperature region, which is similar to that observed in our previous work\cite{DengQPRB} and some 122-type iron-based superconductors\cite{Fisher1,Prozorov1,Uchida,Ying1,Ying2,Prozorov2}. The temperature at which $\bar{R}_a$ and $\bar{R}_b$ start to deviate from each other is defined as $T_{\text{nem}}$. According to the criterion defined in our previous work\cite{DengQPRB}, $T_{\text{nem}}$ is determined on the temperature dependence of $\bar{R}_b-\bar{R}_a$ curve (not shown here). In this case, $T_{\text{nem}}$ is estimated to be $71\pm 5$ K in this study, which is well consistent with our previous report\cite{DengQPRB}.

Figure~2(a) and (b) show the Hall resistivity $\rho_{xy}$ measured when the current is along $a$-axis and $b$-axis, respectively. The magnetic field dependence of Hall resistivity was measured at different temperatures up to 250 K, but the raw data above 60 K were not shown here because the Hall resistivity becomes very small. As shown in Figs.~2(a) and (b), a nonlinear Hall resistivity versus magnetic field can be observed below about 40 K, which is around the antiferromagnetic transition temperature $T_{\text{AF}}$. Make a comparison between Figs.~2(a) and 2(b), one can see that the Hall resistivity under these two configurations are very close to each other (with the difference of less than 3\%), which indicates a similar Hall coefficient between these two configurations.

\begin{figure}
\includegraphics[width=9cm]{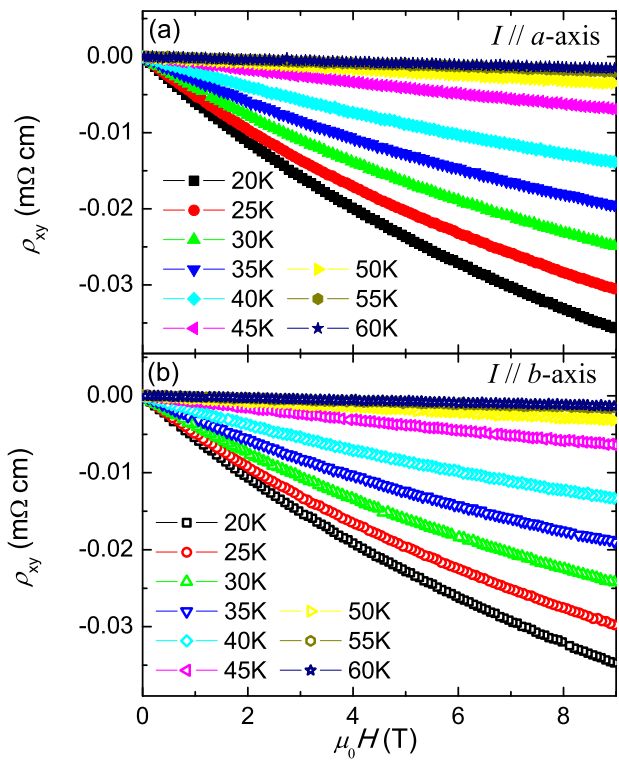}
\caption{Raw data of the Hall resistivity $\rho_{xy}$ measured when (a) $I\parallel a$-axis and (b) $I\parallel b$-axis at temperatures from 20 K to 60 K, with the uniaxial pressure along $b$-axis. It is clear that the Hall resistivity under these two configurations are very close to each other. A nonlinear Hall resistivity versus magnetic field has been observed below about 40 K. }
\end{figure}

Temperature dependence of the Hall coefficient $R_\mathrm{H}$ for two different configurations $I\parallel a$-axis and $I\parallel b$-axis are shown in Fig.~3. $R_\mathrm{H}$ is determined from the slope of $\rho_{xy}$ in the low magnetic field region where the Hall resistivity can be roughly regarded as a linear dependence of magnetic field. The magnitude of $R_\mathrm{H}$ obtained in this study is well consistent with an earlier report in Ref. \cite{GFChen}. The negative value of $R_\mathrm{H}$ over the whole temperature region reveals that the conduction is dominated by electron-like charge carriers. Recall the resistivity data we mentioned above, a clear anisotropy between $\bar{R}_a$ and $\bar{R}_b$ can be observed. In sharp contrast, the Hall coefficient shows a negligible difference under these two configurations. Using the crossing point as shown in the inset of Fig.~3, the Hall coefficient $R_\mathrm{H}$ suddenly increases at a temperature of about 56 K, which is close to the determined structural transition temperature $T_s$.

\begin{figure}
\includegraphics[width=9cm]{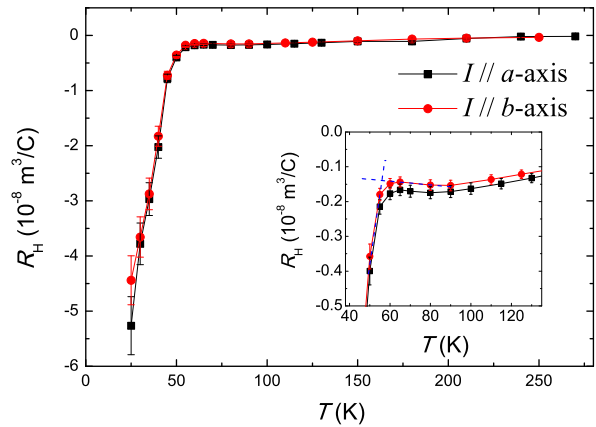}
\caption{Hall coefficient determined through $R_\mathrm{H}=\mathrm{d}\rho_{xy}/\mathrm{d}H$ in the low magnetic field limit with a uniaxial pressure along $b$-axis. For the two different configurations $I\parallel a$-axis and $I\parallel b$-axis, the Hall coefficient shows a very similar temperature dependence, in sharp contrast with the resistivity. The tiny difference between the two set of data of $R_\mathrm{H}(T)$ as shown in the inset were induced by the uncertainty in measuring the size of the electrode silver paste spots for the Hall voltage.}
\end{figure}

Figure~4 shows the temperature dependence of the normalized resistance for the two measuring configurations under magnetic fields from 0 to 9 T. A significant anisotropy of in-plane resistance can be observed below $T_{\text{s}}$ as mentioned above. In addition, a remarkable enhancement of magnetoresistance can be observed below antiferromagnetic transition temperature $T_{\text{AF}}$. Although the resistivity shows the large anisotropy for the two configurations, the magenetoresistance $[R(B)-R(0\;\mathrm{T})]/R(0\;\mathrm{T})$ seems very similar. Take the values at 35 K for example, the ratio of the normalized resistances in the two configurations is about 1.4 while the ratio of the magenetoresistances is 1.08.
Another interesting observation is that below $T_{\text{AF}}$, $\bar{R}_a$ and $\bar{R}_b$ at the same magnetic field decrease in almost parallel way with each other. In other words, $\bar{R}_a$ and $\bar{R}_b$ show quite similar temperature dependent behavior below $T_{\text{AF}}$ under the same magnetic field.

\begin{figure}
\includegraphics[width=9cm]{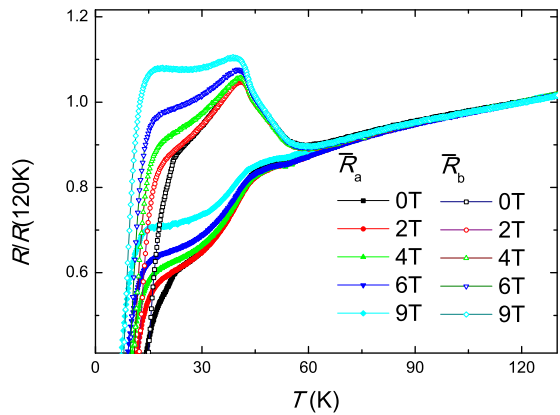}
\caption{Temperature dependence of the normalized resistance with the measuring configurations $I\parallel a$-axis (filled symbols) and $I\parallel b$-axis (open symbols) under magnetic fields from zero to 9 T. There is a significant anisotropy of in-plane resistance under the two different configurations mentioned above.}
\end{figure}

Figures~5(a) and 5(b) show the field dependence of magnetoresistance $\Delta\rho_{xx}/\rho_{xx}(0)$ under two measuring configurations $I\parallel a$-axis and $I\parallel b$-axis, respectively. Here $\Delta\rho_{xx}/\rho_{xx}(0)=(\rho_{xx}(B)-\rho_{xx}(0\;\text{T}))/\rho_{xx}(0\;\text{T})$. As shown in Figs.~5(a) and (b), a large magnetoresistance can be observed below about 50 K. Obviously, the magnetoresistance under these two configurations are close to each other, consistent with the results mentioned above. According to the Kohler's rule, if only one isotropic scattering time $\tau$ dominates in the transport property, $\Delta\rho_{xx}/\rho_{xx}(0)$ should be a function of $H/\rho_{xx}(0)$, then in a Kohler plot $\Delta\rho/\rho_{xx}(0)$ versus $H/\rho_{xx}(0)$, the magnetoresistance data measured at different temperatures should be scalable to one curve\cite{Kohler}. However, as shown in Figs.~5(c) and (d), the data cannot be scaled to one curve at all, the Kohler's rule is severely violated. We will try to understand this discrepancy with the multi-band effect in this material.

\begin{figure}
\includegraphics[width=9cm]{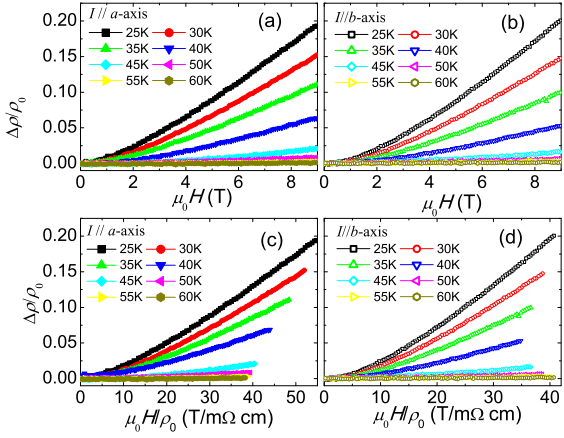}
\caption{Raw data of the magnetoresistance under the two measuring configurations (a) $I\parallel a$-axis and (b)$I\parallel b$-axis. The scaling according to Kohler's rule is given in (c) and (d) with the data shown in (a) and (b), respectively. One can see that the Kohler's rule is severely violated.}
\end{figure}

\section{Analysis on the multiband effect}
From the electric transport measurements, we found that the resistive curves deviate from each other below $T_\mathrm{nem}$ when the current is parallel or perpendicular to the $b$-axis in the detwinned sample. In sharp contrast, the Hall resistivity is almost isotropic under the two different configurations. Furthermore, a non-linear Hall effect as well as a sizeable magnetoresistance are observed when the temperature is below $T_\mathrm{s}$, which is accompanied by the appearance of nematic electronic state. It is not easy to coherently understand the data. The Onsager's theorem would suggest that the Hall effect is isotropic when the scattering rate takes a constant across the Fermi surface. It was argued that the Hall coefficient might be isotropic even with an arbitrary Fermi surface shape\cite{Ong}. While it may not be able to carry out a non-linear Hall effect, nor the sizable magnetoresistance if no magnetic scattering is involved. Furthermore, Even within the Onsager's theorem for one band model, it is unclear that whether the Hall effect is still isotropic if an anisotropic mobility or scattering rate is involved. In addition, the violation of Kohler's rule suggests that the multi-band effect may dominate the electric conductance. From the measurements of ARPES, there are four bands across the Fermi energy, the degeneracy of the $d_{xz}$ and $d_{yz}$ band is lifted in the nematic sate in the detwinned sample\cite{ZXShen,ARPES-Feng}. Transport properties seem to be complex in a multiband system because the contributions of each band entangle each other and give a total conductivity tensor\cite{MgB2-Yang}. Since the band structure is quite complex in this system, we use a two-band picture to investigate this problem quantitatively by assuming that the difference of the two measuring configurations would come from the two main contributions\cite{ARPES-Feng}, such as $d_{xy}+d_{xz}$ and $d_{xy}+d_{yz}$ with different electron scattering affected by the electron-phonon coupling and the impurity scattering. As presented in APPENDIX B, the longitudinal and the transverse resistivity at a magnetic field based on the semiclassical Boltzmann theory with the relaxation time approximation is derived and can be simply expressed as
\begin{figure}
\includegraphics[width=9cm]{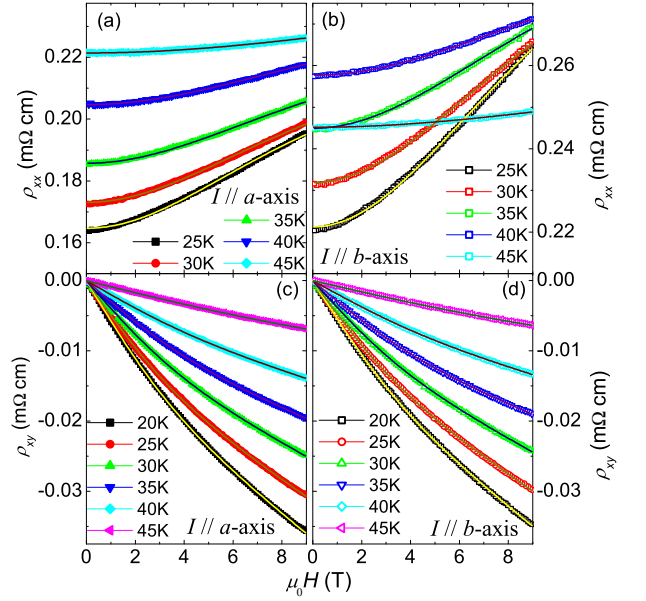}
\caption{Field dependence of longitudinal and transverse resistivity (symbols) and the corresponding theoretical fitting results (solid lines) by using Eqs.~(1) and (2) of two-band model.}
\end{figure}

\begin{equation}
\rho_{xx}(B)=\rho_{xx}(0)\left(1+\frac{P_1B^2}{1+{P_2}^2B^2}\right),
\end{equation}
\begin{equation}
\rho_{xy}(B)=R_\mathrm{H}(0)\left(1+\frac{Q_1B^2}{1+{Q_2}^2B^2}\right)B.
\end{equation}
Here $P_1$, $P_2$, $Q_1$ and $Q_2$ are fitting parameters and $P_2=Q_2$ in a two-band system. In NaFeAs, the measured $\rho_{xy}$ has almost the same field dependent behavior when $I$ is along $a$- or $b$-axis. Then we use Eqs.~(1) and (2) to fit the experimental data of the longitudinal and transverse resistivity with different current directions, and the fitting results are shown as solid lines in Fig.~6. It seems that the two-band model works very well to describe the experimental data. We must mention that the fitting becomes less reliable at high temperatures as the nonlinearity of the magneto-resistivity or the nonlinear Hall effect become weaker, so we only show the fitting parameters for the temperatures below 45 K in Fig.~7. All the fitting parameters, including $R_H(0),P_1,P_2,Q_1,Q_2$ seem to have very little difference between the two configurations, except for the longitudinal resistivity at zero magnetic field ($\rho_{xx}(0)$) which is in agreement with the difference from the original data.

\begin{figure}
\includegraphics[width=9cm]{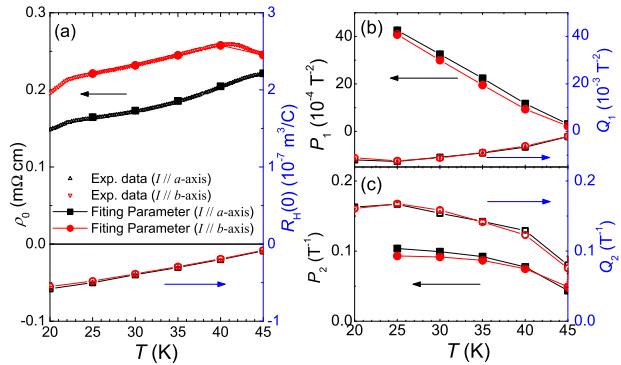}
\caption{Fitting parameters of the experimental longitudinal and transverse resistivity using Eqs.~(1) and (2). The resistivity at zero magnetic field shows an obvious anisotropy when $I\parallel a$-axis and $I\parallel b$-axis, while all other fitting parameters seem to have very little difference under the two configurations.}
\end{figure}

It should be noted that such analysis is based on the data taken at temperatures below $T_\mathrm{s}$, and in this range the magnetoresistance and the non-linear Hall effect are clear enough to investigate the different contribution from the two bands. It seems that the two band model can fit the Hall resistivity and magnetoresistance very well, indicating that both the non-linear Hall resistivity and the strong magnetoresistance are induced by the multiband effect. This is very similar to the multiband effect in MgB$_2$\cite{MgB2-Yang}. In the following we give a deeper insight based on a logical consideration. From the analysis in APPENDIX B, the charge carrier density of each band can be argued to be isotropic according to the experimental observation of isotropic transverse resistivity and anisotropic longitudinal resistivity, while the mobility of the two bands are anisotropic. In this case, it is clear that it is the mobility that governs the strong anisotropic in-plane resistivity. This conclusion is qualitatively consistent with the one-band model where $R_\mathrm{H}$ is equal to $1/ne$, and $\rho_{xx}=1/(ne\mu)$. Therefore the anisotropic in-plane resistivity in the nematic state is related to the mobility. Worthy to mention is that, in principle, the fitting parameters $P_2$ and $Q_2$ should be equal to each other in the two-band model (see Eqs. (11)-(13) in APPENDIX B), but after the values are obtained from fittings to Eq.(1) and(2) respectively, we find that $P_2$ and $Q_2$ have about 40\% difference as exhibited in Fig.~7(c). We don't know what is the detailed reason for this discrepancy. I might suggest that the two-band model is still too simple to catch up the whole physics concerning the non-linear Hall effect and magnetoresistance. However, for the two different configurations, $P_2$ and $Q_2$ are close to each other, therefore the argument mentioned above is still valid.

\section{Discussion}
Our experiments clearly show that the longitudinal resistivity $\rho_{xx}$ becomes anisotropic below the nematic temperature $T_{nem}$. However, the transverse resistivity $\rho_{xy}$ and the Hall coefficient are isotropic in the nematic state. After detailed analysis, as presented in APPENDIX B, we conclude that the strong anisotropic in-plane resistivity is related to the composed mobilities $n_1\mu_{1,j}+n_2\mu_{2,j}$, here $n_{i}$ ($i$=1,2) and $\mu_{i,j}$ denote the charge carrier density and the mobility of the  $i^{th}$ band when they are moving in the $j$-direction ($j=x,y$). Our logical consideration tells that $n_1$ and $n_2$ will not depend on the current direction, but the mobility of each band does. Therefore it is the anisotropic mobility of each band that leads to the clear in-plane anisotropic resistivity. To be precise, as shown by Eq.(7) in APPENDIX B, it is the difference between $n_1\mu_{1a}+n_2\mu_{2a}$ and $n_1\mu_{1b}+n_2\mu_{2b}$ that gives rise to the significant in-plane resistivity. This is qualitatively consistent with the previous results that post-annealing may give strong influence on the anisotropic in-plane resistivity in the nematic state\cite{Nakajima} since the annealing changes either the number and/or the potential of the scattering centers. Actually, the ARPES data\cite{ZXShen} reveal that the degeneracy of the $d_{xz}$ and $d_{yz}$ orbitals is lifted in the nematic state, this naturally leads to a set of Fermi surfaces with a $C_2$ nature, and thus induces anisotropy of Fermi velocity and scattering rate. The in-plane anisotropic resistivity as well as the non-linear Hall effect together with an anisotropic transverse resistivity were observed in an organic superconductor $\kappa$-(BEDT-TTF)$_2$Cu[N(CN)$_2$]Br above about 30 K when a new Fermi surface sheet appears\cite{Tanatar}. The authors describe this as the strong deviation
from the predicted weak-field behavior\cite{Ong}. Interestingly, in the nematic state of iron based superconductors, it was discovered that the antiferromagnetic correlation is established along the $a$-axis after a uniaxial pressure is applied along $b$-axis. From our data shown in Fig.~1, it is clear that the resistivity along the AF direction ($a$-axis) is smaller than that along the direction with parallel spin alignment, the so-called ferromagnetic direction ($b$-axis), this suggests that the resistivity is not induced by the spin scattering effect. This reminds us that, in the pseudogap region of cuprate superconductors, the stripe phase is formed with probably the anisotropic scattering along the two different orthogonal directions. Recently, anisotropic charge dynamics in detwinnedBa(Fe$_{1-x}$Co$_x$)$_2$As$_2$ samples have been observed\cite{Dusza}, which shows difference of the scattering rate and the Drude weight when the polarized light is aligned along the two orthogonal directions. Since the effective Drude weight is also influenced by the effective mass $m^*$, therefore this experiment gives partial support to our results and conclusion. Our results here are calling for more angle resolved spectroscopy measurements that to pin down whether the dramatic in-plane anisotropic resistivity is purely induced by the different mobilities along the two orthogonal directions.

\section{Conclusions}
We measured the longitudinal and transverse resistivity of a NaFeAs single crystal with the configurations: $I\parallel a$-axis and $I\parallel b$-axis when a uniaxial pressure is applied along $b$-axis. The temperature dependence of longitudinal resistivity $\rho_{xx}$ is very different in the two configurations below the structural transition temperature $T_s$, however the transverse resistivity $\rho_{xy}$ and Hall coefficient show almost an isotropic behavior. Large magneto-resistance and non-linear Hall effect are also observed below $T_s$ and the Kohler's rule is severely violated, which suggests the multiband nature in the nematic state. Two-band model with different charge carrier density and mobilities is used to analyze the non-linear Hall effect and the magnetoresistance between the two configurations. Detailed analysis indicates that the moving charge carrier densities $n_1$ and $n_2$ should be isotropic whatever the current direction is, however, there is a clear difference of the composed mobility $n_1\mu_{1a}+n_2\mu_{2a}$ and $n_1\mu_{1b}+n_2\mu_{2b}$, which gives rise to the puzzling and dramatic in-plane anisotropic resistivity in the nematic state.  The present work will stimulate the investigation on the origin of the electronic nematicity in iron based superconductors.

\section*{Acknowledgments}
We thank Makari Tanatar, Zhongyi Lu and Jiangping Hu for helpful discussions and suggestions. We appreciate the kind help from Lei Shan and Xinye Lu in establishing the uniaxial pressure measurement setup. This work was supported by NSF of China, the Ministry of Science and Technology of China (973 projects: 2011CBA00102, 2012CB821403) and PAPD.

\section*{APPENDIX A: MAGNETORESISTANCE AND ITS ANISOTROPY DEGREE}

When the current was applied in different directions, the transverse resistivity has almost the same field dependent behavior. As shown in Fig.~8(b), the value difference of $\rho_{xy}$ at $\mu_0H=9$ T is about 3\% which is within the acceptable error range of the transport measurements. However, the difference ratio of magnetoresistance vary from -3.8\% to 22.2\%, which is obviously beyond the error range of the transport measurements. Thus, magnetoresistance is regarded as anisotropic.

\section*{APPENDIX B: A PROVE OF THE ISOTROPIC CARRIER DENSITY}
We assume that the charge carrier density and the mobility are anisotropic in $x$- or $y$-direction, and use $n_{ij}$ and $\mu_{ij}=e\tau_{ij}/m_{ij}$ as the charge carrier density and mobility of the $i^{th}$ band ($i=1$ or 2) in $j$-direction ($j=x$ or $y$) with $\tau_{ij}$ and $m_{ij}$ the scattering time and effective mass of the $i^{th}$ band in $j$-direction. Based on the semiclassical Boltzmann theory with the relaxation time approximation, the motion equations for the charge carriers of the two-band in the steady state of the system when the current is along $x$-direction of the sample can be described as
\begin{equation}
v_{1x}=\mu_{1x}\left(E_x+v_{1y}B\right), \  v_{1y}=\mu_{1y}\left(E_y-v_{1x}B\right),
\end{equation}
\begin{equation}
v_{2x}=\mu_{2x}\left(E_x+v_{2y}B\right), \  v_{2y}=\mu_{2y}\left(E_y-v_{2x}B\right).
\end{equation}

The net transverse current $J_y=n_{1y}ev_{1y}+n_{2y}ev_{2y}$ must be zero while $\rho_{xx}=E_x/J_x$ and $\rho_{xy}=E_y/J_x$ with $J_x=n_{1x}ev_{1x}+n_{2x}ev_{2x}$. In this situation, Onsager relation is violated, i.e., $\sigma_{yx}(B)\neq-\sigma_{xy}(B)$. Then the longitudinal and transverse resistivity in a system with anisotropic charge carrier density and mobility can be expressed as following,
\begin{widetext}
\begin{equation}
\rho_{xx}(B)=\frac{E_x}{J_x}=\frac{1}{e} \frac{n_{1y}\mu_{1y}+n_{2y}\mu_{2y}+(n_{1y}\mu_{2x}+n_{2y}\mu_{1x})\mu_{1y}\mu_{2y}B^2} {(n_{1x}\mu_{1x}+n_{2x}\mu_{2x})(n_{1y}\mu_{1y}+n_{2y}\mu_{2y})+(n_{1x}+n_{2x})(n_{1y}+n_{2y})\mu_{1x}\mu_{1y}\mu_{2x}\mu_{2y}B^2},
\end{equation}
\begin{equation}
\rho_{xy}(B)=\frac{E_y}{J_x}=\frac{1}{e} \frac{n_{1y}\mu_{1x}\mu_{1y}+n_{2y}\mu_{2x}\mu_{2y}+(n_{1y}+n_{2y})\mu_{1x}\mu_{1y}\mu_{2x}\mu_{2y}B^2} {(n_{1x}\mu_{1x}+n_{2x}\mu_{2x})(n_{1y}\mu_{1y}+n_{2y}\mu_{2y})+(n_{1x}+n_{2x})(n_{1y}+n_{2y})\mu_{1x}\mu_{1y}\mu_{2x}\mu_{2y}B^2}B.
\end{equation}
\end{widetext}
Resistivity and Hall coefficient at $B=0$ read
\begin{equation}
\rho_{xx}(0)=\frac{1}{n_{1x}e\mu_{1x}+n_{2x}e\mu_{2x}},
\end{equation}
\begin{equation}
R_\mathrm{H}(0)=\frac{1}{e}\frac{n_{1y}\mu_{1x}\mu_{1y}+n_{2y}\mu_{2x}\mu_{2y}}{(n_{1x}\mu_{1x}+n_{2x}\mu_{2x})(n_{1y}\mu_{1y}+n_{2y}\mu_{2y})}.
\end{equation}
According to the Eqs.~(5)-(8), we can obtain the simplified expression of the longitudinal and transverse resistivity as Eqs.~(1) and (2). In the situation of NaFeAs, when the current is along $a$- or $b$-axis, the transverse resistivity can be written as
\begin{widetext}
\begin{equation}
\rho_{xy}^{I\parallel a}=\frac{1}{e} \frac{n_{1b}\mu_{1a}\mu_{1b}+n_{2b}\mu_{2a}\mu_{2b}+(n_{1b}+n_{2b})\mu_{1a}\mu_{1b}\mu_{2a}\mu_{2b}B^2} {(n_{1a}\mu_{1a}+n_{2a}\mu_{2a})(n_{1b}\mu_{1b}+n_{2b}\mu_{2b})+(n_{1a}+n_{2a})(n_{1b}+n_{2b})\mu_{1a}\mu_{1b}\mu_{2a}\mu_{2b}B^2}B,
\end{equation}
\begin{equation}
\rho_{xy}^{I\parallel b}=\frac{1}{e} \frac{n_{1a}\mu_{1a}\mu_{1b}+n_{2a}\mu_{2a}\mu_{2b}+(n_{1a}+n_{2a})\mu_{1a}\mu_{1b}\mu_{2a}\mu_{2b}B^2} {(n_{1a}\mu_{1a}+n_{2a}\mu_{2a})(n_{1b}\mu_{1b}+n_{2b}\mu_{2b})+(n_{1a}+n_{2a})(n_{1b}+n_{2b})\mu_{1a}\mu_{1b}\mu_{2a}\mu_{2b}B^2}B.
\end{equation}
\end{widetext}

\begin{figure}
\includegraphics[width=8cm]{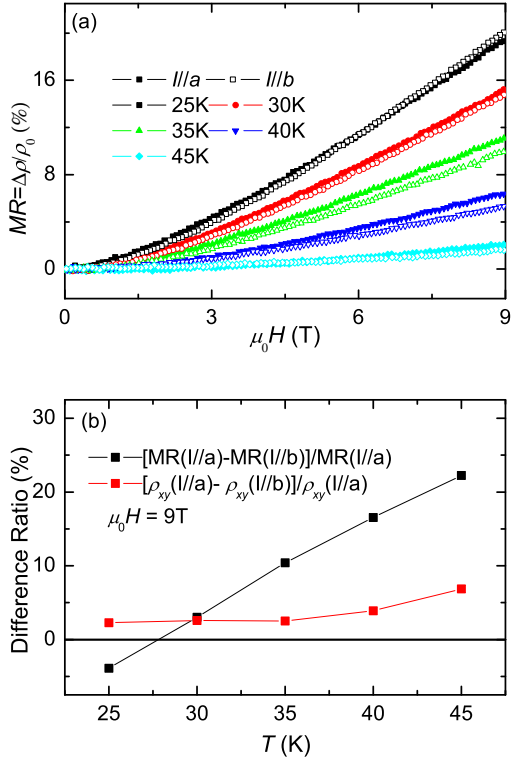}
\caption{(a). Field dependence of magnetoresistance at different temperatures when $I\parallel a$-axis and $I\parallel b$-axis. (b). Temperature-dependent difference ratio of transverse resistance and magnetoresistance at $\mu_0H=9$ T.}
\end{figure}

Since the transverse resistivity has the same magnetic field dependent behavior as $\rho_{xy}^{I\parallel a}(B) = \rho_{xy}^{I\parallel b}(B)$ when the current is along $a$- or $b$- direction, so the coefficients of Eqs.~(9) and (10) on the numerator should be the same. Then we get two possible solutions: (1) $n_{1a}=n_{1b}$ and $n_{2a}=n_{2b}$, or (2) $\mu_{1a}\mu_{1b}=\mu_{2a}\mu_{2b}$ and $n_{1a}+n_{2a}=n_{1b}+n_{2b}$. In the same model, the magnetoresistance ($MR$) can be described as following,
\begin{widetext}
\begin{equation}
MR^{I\parallel a}(B)= \frac{\mu_{1b}\mu_{2b}(\mu_{1a}-\mu_{2a})(n_{1a}n_{2b}\mu_{1a}-n_{1b}n_{2a}\mu_{2a})B^2} {(n_{1a}\mu_{1a}+n_{2a}\mu_{2a})(n_{1b}\mu_{1b}+n_{2b}\mu_{2b})+(n_{1a}+n_{2a})(n_{1b}+n_{2b})\mu_{1a}\mu_{1b}\mu_{2a}\mu_{2b}B^2},
\end{equation}
\begin{equation}
MR^{I\parallel b}(B)= \frac{\mu_{1a}\mu_{2a}(\mu_{1b}-\mu_{2b})(n_{1b}n_{2a}\mu_{1b}-n_{1a}n_{2b}\mu_{2b})B^2} {(n_{1a}\mu_{1a}+n_{2a}\mu_{2a})(n_{1b}\mu_{1b}+n_{2b}\mu_{2b})+(n_{1a}+n_{2a})(n_{1b}+n_{2b})\mu_{1a}\mu_{1b}\mu_{2a}\mu_{2b}B^2}.
\end{equation}
\end{widetext}
If we apply one resultant $\mu_{1a}\mu_{1b}=\mu_{2a}\mu_{2b}$ from the isotropic field-dependent transverse resistivity to above two formulas, we will also obtain $MR^{I\parallel a}(B)=MR^{I\parallel b}(B)$, which is inconsistent with the experimental results as illustrated in APPENDIX A. In this case, the only conclusion from the isotropic transverse resistivity is the isotropic charge carrier density in this system, i.e., $n_{1a}=n_{1b}$ and $n_{2a}=n_{2b}$. This naturally grantees the isotropic field dependent transverse resistivity.
\begin{widetext}
\begin{equation}
\rho_{xy}^{I\parallel a}=\rho_{xy}^{I\parallel b}=\frac{1}{e} \frac{n_1\mu_{1a}\mu_{1b}+n_2\mu_{2a}\mu_{2b}+(n_1+n_2)\mu_{1a}\mu_{1b}\mu_{2a}\mu_{2b}B^2} {(n_1\mu_{1a}+n_2\mu_{2a})(n_1\mu_{1b}+n_2\mu_{2b})+(n_1+n_2)(n_1+n_2)\mu_{1a}\mu_{1b}\mu_{2a}\mu_{2b}B^2}B.
\end{equation}
\end{widetext}

\end{document}